# Demonstration of a quantized acoustic octupole topological insulator

Xiang Ni[1,2,3], Mengyao Li[1,2], Matthew Weiner[1,2], Andrea Alù[3,2,1], and Alexander B. Khanikaev[1,2]

[1]Department of Electrical Engineering, Grove School of Engineering, City College of the City University of New York, 140th Street and Convent Avenue, New York, NY 10031, USA.

[2]Physics Program, Graduate Center of the City University of New York, New York, NY 10016, USA.

[3]Photonics Initiative, Advanced Science Research Center, City University of New York, New York, NY 10031, USA

**Recently extended from the modern theory of electric polarization[1-4], quantized multipole topological insulators (QMTIs)[5,6] describe higher-order multipole moments, lying in *nested* Wilson loops, which are inherently quantized by lattice symmetries. Overlooked in the past, QMTIs reveal new types of gapped boundaries, which themselves represent lower-dimensional topological phases and host topologically protected zero-dimensional (0D) corner states. Inspired by these pioneering theoretical predictions, tremendous efforts have been devoted to the experimental observation of topological quantized quadrupole phase in a variety of two dimensional (2D) metamaterials[7-10]. However, due to stringent requirements of anti-commuting reflection symmetries in crystals, it has been challenging to achieve higher-order quantized multipole moments, such as octupole moments, in a realistic three-dimensional (3D) structure. Here, we overcome these challenges, and experimentally realize the acoustic analogue of a quantized octupole topological insulator (QOTIs) using negatively coupled resonators. We confirm by first-principle studies that our design possesses a quantized octupole topological phase, and experimentally demonstrate spectroscopic evidence of a topological hierarchy of states in our metamaterial, observing 3$^{rd}$ order corner states, 2$^{nd}$ order hinge states and 1$^{st}$ order surface states. Furthermore, we reveal topological phase transitions from higher- to lower-order multipole moments in altered designs of acoustic TIs. Our work offers a new pathway to explore higher-order topological states (HOTSs) in 3D classical platforms.**

Since the discovery of the quantum Hall effect,[11] topological phases of matter have attracted a significant attention due to their unique physical properties. While most interest in past years has been driven by research in condensed matter physics[12-15], classical systems have proven as powerful and versatile platforms to realize a wide variety of topological phases[16-20]. The inherent advantage in realizing topological phases in classical systems stems from the possibility of implementing a wide range of artificial potentials and gauge fields acting on engineered synthetic degrees of freedom. Thus, classical systems not only offer the possibility of emulating known quantum topological phenomena, but also allow to test new topological phases that may be hard or even impossible to find in naturally occurring materials. This is especially important in the context of recent predictions of a variety of novel topological phases of matter, which can be immediately tested using designer acoustic or photonic materials. One of such recently predicted topological phases, which is of significant theoretical and practical interest, and is at the core of this paper, is the class of QMTIs [5,6].

The concept of electric multipole moment has been introduced by Benalcazar et al.[5,6] as a generalization of electric polarization (dipole moment), which is essential to understand the behavior of foundational topological phenomena such as polarization-induced boundary charge in polyacetylene chains[21] and edge currents in quantum Hall systems[22-24]. In contrast to previously studied topological systems, whose topological phase arises from bulk energy bands and the associated 1st order Wannier band, the topology of QMTIs emerges from a hierarchy of gapped bulk, surface, and hinge bands, and associated nested (higher-order) Wannier bands. The unique property of such multipole TIs is manifested in HOTSs localized not only at the surface, but also at edges and corners of the system. In fact, the emergence of quantized charge corner states represents a hallmark of multipole TIs. While corner, edge and surface states can also be found in systems with topological bulk polarization[25-33], as well as the second order modes protected by crystalline symmetries were reported in the pioneering work by Noh et al.[34] (and more recently studies in nonlinear regimes [35]), it is only in QMTIs that higher multipole moments are enforced by the lattice symmetries to be quantized to 0 or 1/2. This property has been experimentally confirmed in 2D microwave, mechanical and circuit based systems.[7-10] However, the extension to 3D quantized octupole TIs is not straightforward for multiple reasons: first, because of the requirement that octupole TIs must provide a synthetic magnetic flux of $\pi$ through every plaquette of a 3D crystal, second, because they must possess dimerized inter-cell and intra-cell hopping in all three dimensions, and third, because, at the same time, they must preserve reflection symmetries to protect the topological phase. Here, using a versatile additive manufacturing platform, we successfully tackle all the technical obstacles and experimentally implement these requirements in a 3D acoustic metamaterial, demonstrating for the first time an acoustic TI with quantized octupole moment supporting a hierarchy of topological states, including third-order corner states. Besides experimentally demonstrating the spectral evidence of non-trivial octupole moment, we also show through first principle simulation the rich topological classes that can be supported by the proposed design.

A 3D acoustic octupole TI proposed here emulates the tight-binding model (TBM) shown in Fig. 1**a**, which was originally introduced in refs[5,6]. The model is emulated using unit cells consisting of eight coupled acoustic resonators of cylindrical shape, as schematically shown in Fig. 1**b**. We choose to work with the lowest-order axial acoustic mode, whose pressure oscillates along the cylinder (see Fig. 1**d** as an example profile). To form the lattice, the resonators are coupled through narrow cylindrical channels, whose position on the resonator is carefully chosen to respect the crystalline symmetries. A $\pi$ synthetic magnetic flux through each plaquette is achieved by bending the channels connecting the resonators based on the protocol shown in Fig. 1**b**. Assuming the simplest case, the bent channels (blue colored connectors in Fig. 1**b**) provide a zero phase shift in the coupling, while the straight channels (red colored connectors in Fig. 1**b**) connect the bottom of one cylinder with the top of the other cylinder in the *xy*-plane, providing an additional $\pi$ phase shift in the hopping due to the odd symmetry of the modes of interest. More general scenarios are examined in the Extended Data Figure 1, which shows that the protocol implemented here always ensures a synthetic magnetic flux of $\pi$ through each plaquette. In addition, the channel lengths are chosen to be identical in all directions, in order to guarantee no dynamic phase difference in the coupling channels due to propagation. At the same time, the dimerization of intra-cell coupling $\gamma_i$ and inter-cell coupling $\lambda_i$ in all directions is achieved by varying the diameter of the coupling channels, such that $|\gamma_i| < |\lambda_i|$. Since the acoustic modes are strongly bound to the resonators and the design of the structure only allows nearest neighbor coupling, the tight-binding model (TBM) in Fig. 1**a** is nicely reproduced in our acoustic crystal. The only distinction from the original model in Refs.[5,6] is the tetragonal geometry of our lattice, as opposed to the cubic one, which is due to the bending taking place in the *xy*-plane and different intra-cell coupling in the directions parallel and perpendicular to the *xy*-plane. This design-imposed alterations have no consequences on the topological phase of the system, as the essential inversion and reflection symmetries are preserved [5], leading to the vanishing of bulk dipole and quadrupole moments and the quantization of octupole moment. Detailed symmetry analysis of the Hamiltonian constraint in the context of this higher-order topology is presented in the Supplementary Information. COMSOL Multiphysics (Acoustic Module) has been used to verify our results with full-wave finite element method (FEM) simulations, and the bulk acoustic band structure along high-symmetry directions of the tetragonal Brillouin zone (BZ) of the proposed octupole TI is shown in Fig. 1**c**, revealing two sets of four-fold degenerate bands split around the frequency of a single isolated resonator ("zero energy" of our system), spectrally separated from other modes. The symmetry of the band structure with respect to the frequency of a single resonator, and the degeneracy of the bulk bands, further confirm the presence of essential reflection, inversion, and sublattice (chiral) symmetries[5]. The bulk field profile shown in Fig. 1**d** clearly reveals the effect of chosen coupling protocol on the phase distribution of coupling between dipolar modes.

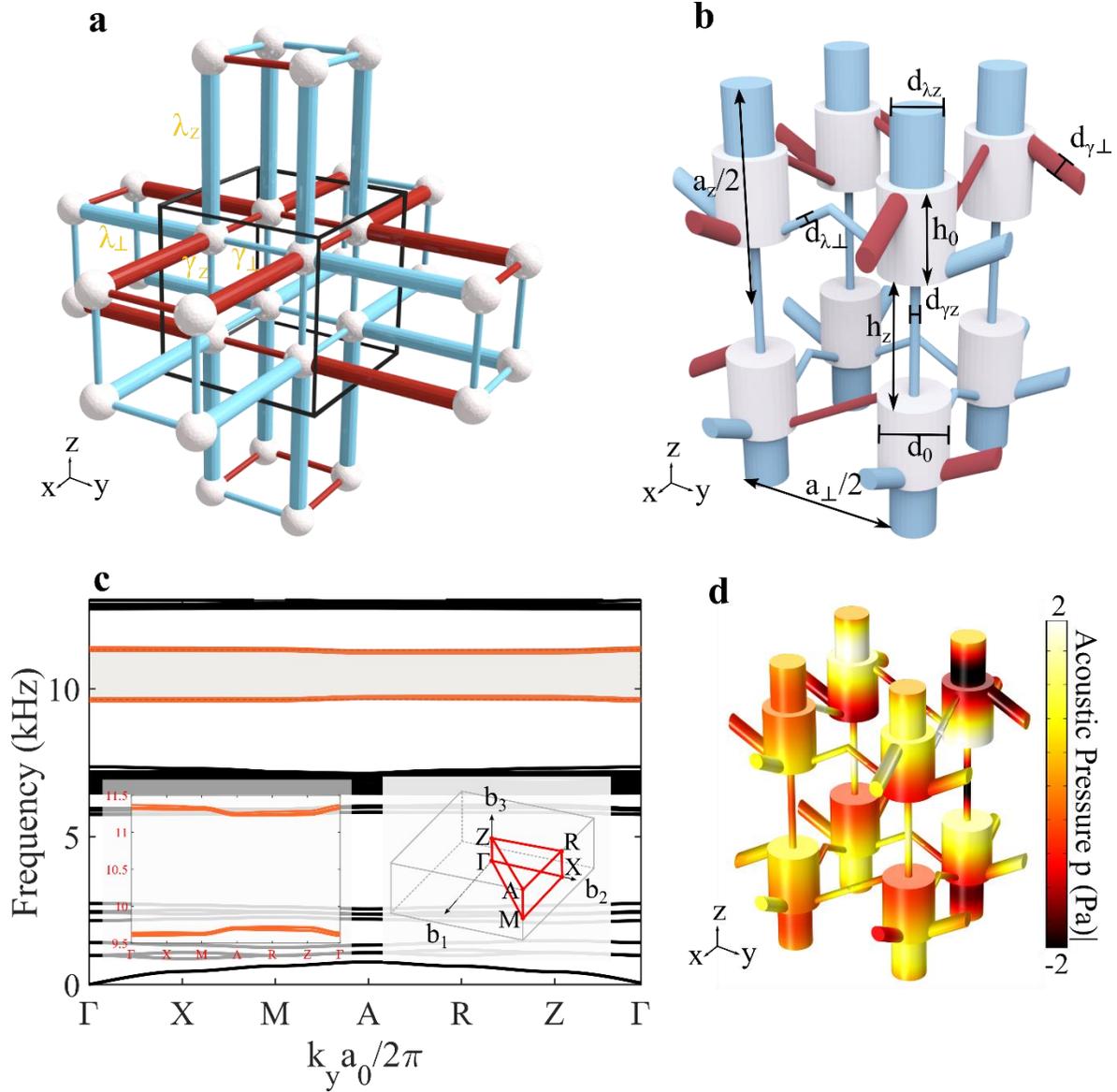

**Fig. 1 | Acoustic octupole topological insulator**. **a**, geometry for minimal tight-binding model of octupole topological insulator, $\gamma_\perp$ and $\lambda_\perp$ are the intra-cell and inter-cell couplings in the plane perpendicular to $z$ axis, and $\gamma_z$ and $\lambda_z$ are the intra-cell and inter-cell couplings in the $z$ direction. **b**, unit cell design of acoustic tetragonal crystal that guarantees the negative coupling at each plaquette of the lattice, geometry parameters are provided in the Methods. **c**, band structures of the acoustic unit cell in **b**, shaded grey region indicates the frequency bandwidth of interested bands (orange colored), inset figures show four-fold degenerate band structures of interest (slight deviation due to meshing in numerical calculation) and high symmetry lines and points in the tetragonal Brillouin zone, respectively. **d**, Acoustic pressure profile of the acoustic coupled dipolar modes taken at the high symmetry point A.

To prove that the proposed acoustic structure indeed possesses quantized octupole topological phase, we carried out calculations of the higher-order Wannier bands based on first-principle FEM simulations. As shown in Fig. 2 **a-c**, the structure supports gapped two-fold degenerate first-order

Wannier bands ($v_z$), gapped second-order Wannier bands ($v_x^{-z}$), and importantly, a third-order Wannier band ($v_y^{-x,-z}$) quantized as $1/2$. The same conclusions are obtained for Wannier bands with arbitrary order in $x, y, z$. Thus, the octupole topological phase (Wannier-sector polarization) of the acoustic tetragonal crystal is

$$\left(p_x^{-z,-y}, p_y^{-x,-z}, p_z^{-y,-x}\right) = \left(\tfrac{1}{2}, \tfrac{1}{2}, \tfrac{1}{2}\right). \tag{1}$$

Further explanations on the appearance of gapped Wannier bands induced by non-commuting reflection symmetries are presented in the Supplementary Information. Since the Wannier bands are adiabatically connected to the energy spectrum of corresponding boundary states[36], surface states and hinge states are expected to be gapped, while the corner states which arise from quantized bulk octupole moment are anticipated to be in the mid-gap of the energy spectrum due to the chiral symmetry (see Supplementary Information for details). We emphasize that it is only the corner states are pinned to "zero energy" and protected by the chiral symmetry, and the vanishing dipole moment and quadrupole moment enforced by inversion symmetries in our system make hinge and surface states trivial. All these predictions are indeed confirmed by first-principle simulations of our structure in Fig. 2**d-f**, which clearly reveal gapped spectra with nested surface (yellow bands), hinge (blue bands), and corner states (red bands). The associated field profiles for (i) the top surfaces of the crystal (Fig. 2**g**), (ii) the corresponding edges of top surfaces (Fig. 2**h**), and (iii) the corners of top surfaces (Fig. 2**i**) verify the boundary nature of these states. Note that, due to the anisotropy of the crystal, the surface and hinge states of vertical boundaries support a wider bandgap (Fig.2**f**, gray-colored bands). In addition, the frequency splitting of corner states (~10Hz) in Fig. 2f is due to finite size of the system and hybridization of the corner modes localized at different corners.

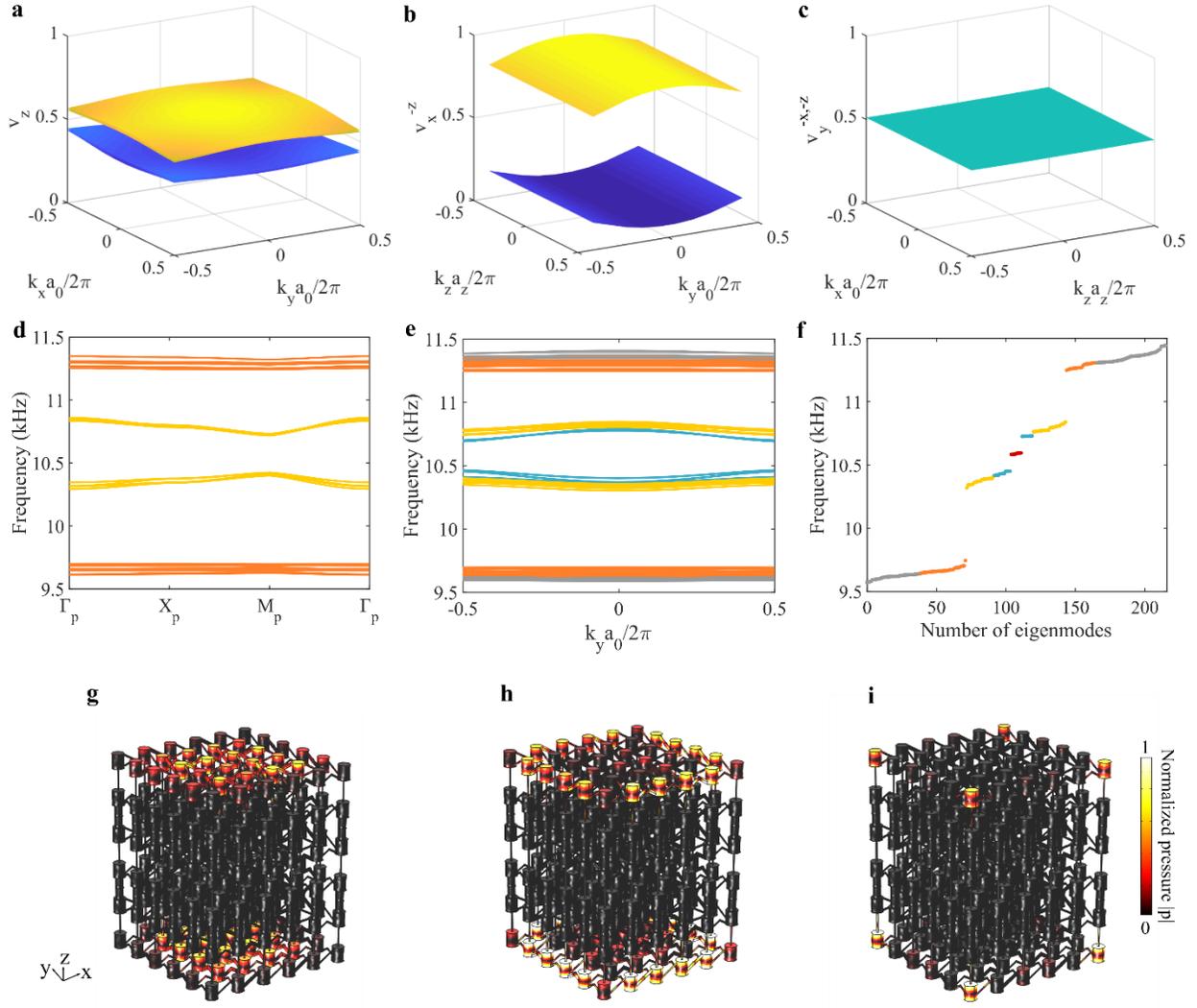

**Fig. 2 | Higher-order topology and corresponding boundary states of the acoustic tetragonal crystal.**
**a-c**, *n*-th order Wannier bands obtained based on the first principle FEM simulation. **a,** 1st order Wannier bands $v_z$ of the selective energy bands below the bandgap shown in Fig. 1 **c**. **b**, 2nd order Wannier bands $v_x^{-z}$ of the Wannier-sector $-v_z$. **c**, 3rd order Wannier bands $v_y^{-x,-z}$ of the Wannier sector $-v_x^{-z}$. **d-f**, energy bands show hierarchy topological states in which **d**, energy bands of the supercell lattice (110) supporting the surface states (yellow colored bands). **e**, energy bands of the supercell lattice (010) supporting the hinge states (blue colored bands). **f**, energy diagram of a finite lattice composed of $3 \times 3 \times 3$ unit cells supporting mid-gap edge polarization induced corner states. **g-i**, surface, hinge and corner field distributions over the finite lattice integrated from the respective boundary eigenstates inside the bulk bandgap in **f**. Surface and hinge states in the vertical boundaries are mixed and represented by grey colored bands in **f**.

The experimental sample implementing our octupole TI was fabricated with the use of a high-resolution stereolithographic (STL) 3D printer (see Methods). The unit cells were printed and snapped together using interlocking features deliberately introduced in the design, thus allowing for the assembly of a rigid and stable large-scale crystal. The assembled structure consists of 27 unit cells (216 resonators in total), as shown in Fig. 3**a**. The modes of the fabricated structure were probed with a local excitation in each resonator by placing an air transducer at the side-hole, intentionally introduced into every resonator. The strength of the local response was measured using a microphone attached to the second side-hole. The holes are small enough not to introduce excessive loss, but sufficiently large to probe the acoustic pressure field. The resultant frequency-response spectra for selected groups of resonators (Fig 3**b** inset), internal bulk, and external surface, hinge, and corner resonators, are shown in Fig. 3**b** by color-coded bands, and clearly reveal four distinct types of states (see Methods for detailed measurement and data analysis). The average quality factor of the modes of resonators is about 90 due to loss in the resin and leakage through probe holes giving the mean resonant bandwidth of about 120Hz. Compared to the gaps of bulk (~1500Hz), surface and hinge states (~400Hz), these resonant peaks are narrow enough to clearly distinguish the corner states in the spectra. As predicted by our theoretical calculations, the corner states appear to be nested within gapped hinge, surface, and bulk spectra. The field profiles at the specific frequencies corresponding to bulk, surface, hinge, and corner states are shown in Fig. 3**c-f**, and confirm the localization of these states to the corresponding boundary resonators. Frequency response spectra for eight corners are shown in Extended Data Fig. 2, providing further evidence of octupole topological phase in our structure. We point out the resonant peaks of corners fluctuate around the mid-gap frequency since 3D printing used in fabricating the sample has resolution of ~70um, which is the main mechanism of the chiral symmetry reduction due to fluctuations of the resonant frequency of the cylinders. Together with other mechanisms, such as non-uniform focusing of light in stereolithographic 3D printer, leads to the resonant frequency variations excededing100Hz. However, we minimize the role of such chiral symmetry reduction by individually testing the resonators to ensure approximately the same resonant frequency, such that the frequency splitting of corner modes stays within 100Hz, which is sufficiently small even with respect to the narrowest (hinge states) band gap (~400Hz), as indicated by the dashed lines in Extended Data Fig. 2.

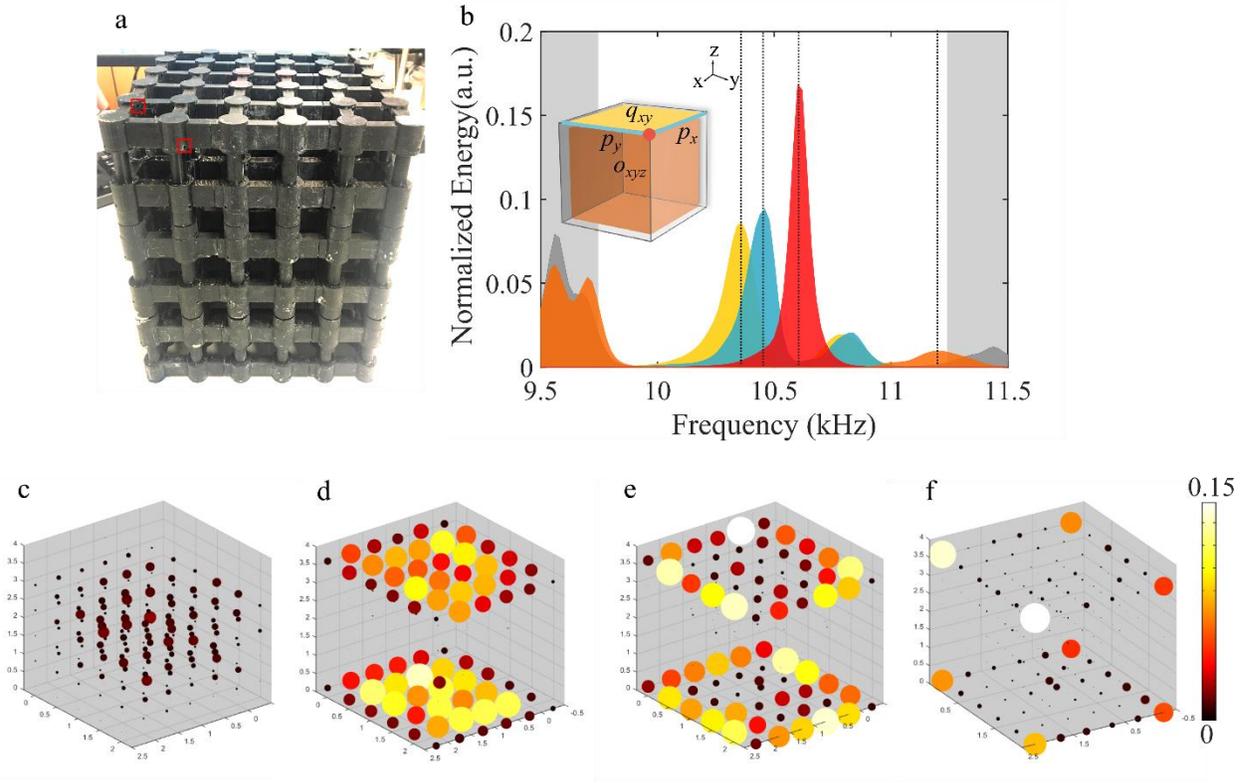

**Fig. 3 | Experimental demonstration of acoustic octupole TI**. **a**, photograph of the assembled structure consisting of $3 \times 3 \times 3$ unit cells. Location of probe holes is shown by red squares. **b**, Normalized acoustic frequency response spectra for the selective bulk (orange), surface (yellow), edge(blue) and corner sites(red) of a finite structure. Grey colored spectra represent surface and hinge states in the vertical boundaries perpendicular yellow colored surface. Inset is the schematic of selective groups of resonators, in which their respective topologies are shown, like bulk has octupole moment $o_{xyz}$, top surface has octupole moment $q_{xy}$, and top edges have dipole moment $p_x$ and $p_y$. **c-f**. The field profiles of **c**, bulk states, **d**, surface states, **e**, hinge states and **f**, corner states extracted from the frequency responses of all resonators and chosen at selected frequencies indicated by dashed lines in **b**.

The insulator with non-commuting reflection symmetries possesses rich topological classes, as illustrated by our first-principle FEM studies summarized in Fig. 4**a**, where the dimerization between $\lambda_i$ and $\gamma_i$ are changed along the selected direction $i$. When the ratio $\left|\frac{\lambda_i}{\gamma_i}\right|$ crosses one, the corresponding 2nd order Wannier bandgap $v_j^{\pm k}$ closes and reopens, implying a topological transition, therefore inferring that the Wannier-sector polarization $p_i^{\pm j, \pm k}$ switches between $1/2$ and $0$. For example, the quantized octupole moment $o_{xyz} = 1/2$ lies in the topological class represented by the red block of Fig. 4**a**. If the dimerization of couplings in the *y* direction changes to $\left|\frac{\lambda_y}{\gamma_y}\right| < 1$, the insulator enters into the topological class of blue block in Fig. 4**a**, corresponding to the scenario of crystal with nontrivial quadrupole phases in the *zx*-plane$\left(p_x^{-z,-y}, p_y^{-x,-z}, p_z^{-y,-x}\right) =$

($\frac{1}{2}$,0,$\frac{1}{2}$), which supports surface and hinge states induced by such quadrupole moment nested in the bulk bandgap in Fig. 4**b**. However, the octupole moment becomes trivial in this class, thus no mid-gap topological corner states are observed. The reversal of dimerization along *y*, *x* and then *z*-direction one by one leads to a sequence of topological transitions, accompanied by a sequence of closing and reopening of 2$^{nd}$ order Wannier bands, corresponding to the disappearance of corner, hinge, and the surface states accordingly, as shown in Fig. 4**b,c,d**.

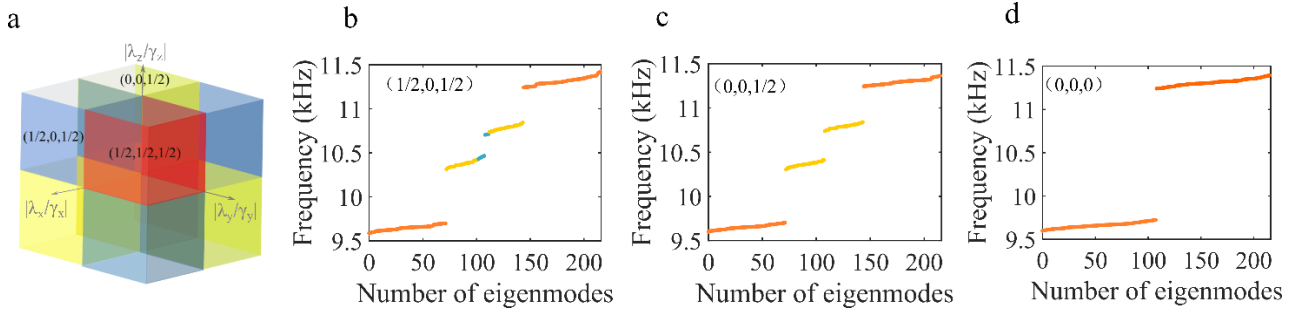

**Fig. 4 | topological classes of the 2$^{nd}$ order Wannier bands.** a, diagram of the topological classes for the 2$^{nd}$ order Wannier bands as a function of $\left|\frac{\lambda_i}{\gamma_i}\right|, i = x, y, z$. **b-d**, energy spectra of a finite lattice obtained from first principle FEM studies with a topological phase **b**, in $(p_x^{-z,-y}, p_y^{-x,-z}, p_z^{-y,-x}) = (\frac{1}{2}, 0, \frac{1}{2})$, **c**, in $(p_x^{-z,-y}, p_y^{-x,-z}, p_z^{-y,-x}) = (0, 0, \frac{1}{2})$, and **d**, in $(p_x^{-z,-y}, p_y^{-x,-z}, p_z^{-y,-x}) = (0, 0, 0)$.

To conclude, in this Letter we reported the design, fabrication and experimental verification of a 3D octupole topological insulator using a classical acoustic platform. We confirmed that the octupole topological moment leads to the emergence of a hierarchy of higher-order boundary states, topological 3$^{nd}$ order states confined to corners, 2$^{nd}$ order states localized at the edges, and 1$^{st}$ order states localized at the surface of the crystal. The corner modes appear in the gap of all lower-order states, making them good candidates for precise control of energy localization in the lattice. The versatile engineering of synthetic magnetic flux and acoustic potentials in 3D enabled by the complex geometries of the 3D printed metamolecules implemented in this work further facilitates design and control of sound based on topological properties. Our results open new avenues in exploring novel ideas of topological physics and suggest new approaches in localizing, guiding, and controlling sound in complex acoustic metamaterials.

We note a recent work, which proposes an alternative interpretation of quantized multipole topological insulators, including octupole HOTI studied here, by considering them as boundary obstructed topological phases. [37].

**Methods**

*Structure design and 3D printing–* The geometries of the unit cell in Fig. 1**b** are as follow: the lattice constants of the tetragonal crystal in *xy*-plane and along *z*-axis are $a_\perp = 67.50 mm$ and $a_z = 81.48$mm, respectively. The height and diameter of the cylinder are chosen as $h_0 = 16.20 mm$ and $d_0 = 13.50$mm, such that the frequencies of the desired modes are in the probing range of the microphone, and also far away from the undesired (transverse) modes. The cylinders are coupled through either bended or straight cylindrical channels, they have the same acoustic length (the dynamic phase is the same) in all directions. The coupling strength of the modes is adjusted by changing the diameter of channels. In order to make the intra-cell and inter-cell coupling of the structure inequivalent, in the octupole TI case, the diameters of the intra-cell channels are set to $d_{\lambda\perp} = 3.78mm$, $d_{\lambda z} = 9.07mm$, while the diameters of the inter-cell channels are set to $d_{\gamma\perp} = d_{\gamma\perp} = 1.89mm$. In the cases of other topological classes, geometrical sizes are the same as above except the diameters of intra-cell and inter-cell channels are exchanged in the selected directions.

The unit cells were fabricated using the B9Creator v1.2 3D printer. All cells were made with acrylic-based light-activated resin, a type of plastic that hardens when exposed to UV light. Hard wall boundary condition are ensured by a sufficient thickness of the printed structure. Narrow probe holes with the diameter $D_0 = 2.00$ mm were intentionally introduced on opposite sides of each of the cylinders to excite and measure local pressure field at each resonator.

*Numerical method* – Finite element solver Multiphysics Comsol 5.2a and the Acoustic module was used to perform full-wave simulation. In the acoustic propagation wave equation, the speed of sound was set as $343.2 \ m/s$, and density of air as $1.225 kg/m^3$. For bulk (surface, hinge) band structure calculations, the Floquet periodic boundary conditions were imposed along the edges of the unit cell (supercell). Large-scale simulations were performed with hard wall boundary conditions applied on the boundaries. To calculate the higher order Wannier bands from simulation results, the eigenstates are constructed by the complex values of acoustic pressure field extracted at the center of top surface of each cylinder for the modes of interest (four-fold degenerate bulk modes below the bandgap).

The on-site frequency (the single resonator frequency, which also plays the role of "zero-energy"), intra-cell and inter-cell couplings are fitted by the TBM with $\omega_0 \cong 10590$Hz, $|\lambda_\perp| \cong 197$Hz, $|\lambda_z| \cong 767$Hz, $|\gamma_\perp| \cong 30$Hz and $|\gamma_z| \cong 42$Hz. The value of "zero energy" is extracted from the resonant frequency of an individual resonator (decoupled from the lattice) via both first-principle (finite element method) FEM numerical simulations and experimental measurements, which agree well. Similarly, the hopping κ between the two resonators connected by a channel in any direction is extracted by fitting the separation between the resonances of the system $\omega_2 - \omega_1 = 2\kappa$. The negative sign of coupling can be also immediately confirmed from FEM studies by inspecting the acoustic field distribution. Subsequently, we verify that the extracted parameters give agreeable results between first-principle the FEM model and the TBM.

*Measurement and signal analysis-* For all measurements, a frequency generator and FFT spectrum analyzer scripted in LabVIEW were used. The speaker was placed at one of the side holes of the cylinder, and the microphone at the other side hole of the same cylinder. A tiny gap was left between the speaker and the port to allow for the presence of reflection channels while the microphone was closely touched with the port to achieve the maximum absorption. The frequency generator was used to run a sweep from 9300 Hz to 11800 Hz in 20 Hz intervals and with the dwell time of 1 seconds while the FFT spectrum analyzer obtained the amplitude responses $\varphi(f)$ at each frequency $f$. Field distributions $\varphi(i, f)$ are obtained by repeating this process for each site $i$. Since it's hard to guarantee the tiny gap is exactly the same for every site, and the amplitude response is highly sensitive to this tiny gap, we normalized the data for each site based on the total squared signal summed over frequencies, $\Phi(i, f) = |\varphi(i, f)|^2 / \sum_f |\varphi(i, f)|^2$. After that, we averaged the energy spectrum for a group of chosen sites (inset of Fig. 3**b**) $P_a(f) = \sum_i \Phi(i, f) / N_o$ to get the normalized spectra for groups of bulk, surface, hinge and corner sites with $N_o = 64, 16, 8, 1$, respectively.

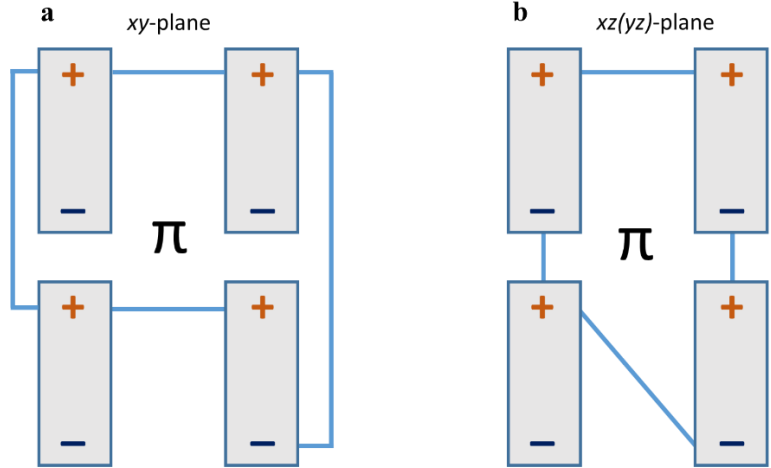

**Extended Data Figure 1 | Flux π through each plaquette a simple case which can be extended to arbitrary case. a,** a simple distribution of acoustic dipole modes in the *xy*-plane of our structure. If we assume interaction between high pressure and lower pressure (+,-) is negative, and interaction (+,+) or (-,-) is positive, then the plaquette has flux π due to the negative coupling. If $n$ number of dipole modes flip their phases, varied flux caused by such change is $2n\pi$, where $n = 1,2,3,4$, thus the total flux π is invariant. **b**, a simple distribution of acoustic dipole modes in the *xz*-plane (*yz*-plane) of our structure. The same analysis can be applied and we get the conclusion that total flux is always π through the plaquette in *xz*-plane (*yz*-plane).

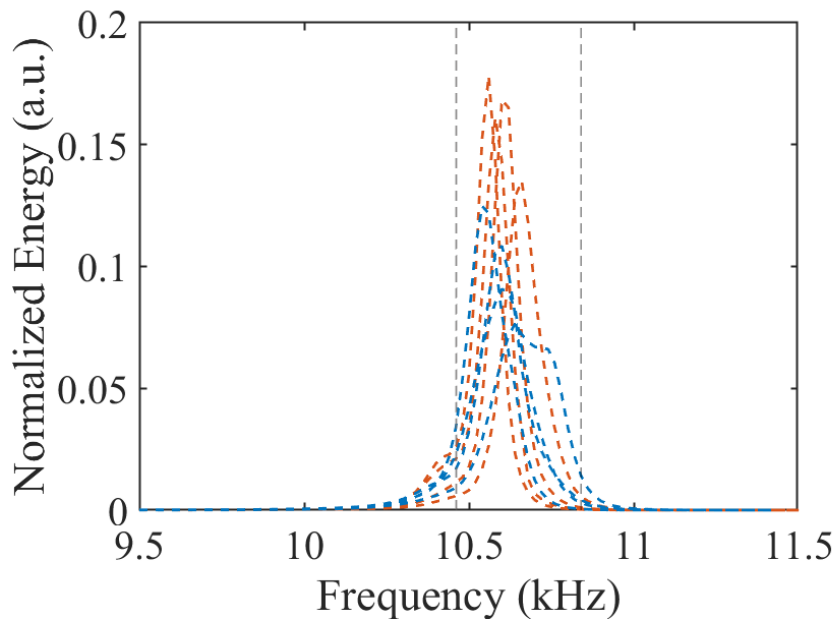

**Extended Data Figure 2 | Frequency responses of single resonator at eight corners, respectively.** Orange and blue colored spectra are for corners at the top and bottom surfaces of the structure in Fig. 3**a,** dashed lines indicate the bandgap range of topological hinge states (~400Hz). The limited resolution of 3D fabrication as well as the loss induced by material absorption and leaky channels lead to the fluctuation of

resonant peaks and non-uniform broadening of corner modes located at different corner sites. The overall higher loss for lower surface are due to purely technical reasons, including larger surface roughness inside the resonators and the deliberate lower print quality (to accelerate fabrication) affecting the internal structure of the resin.

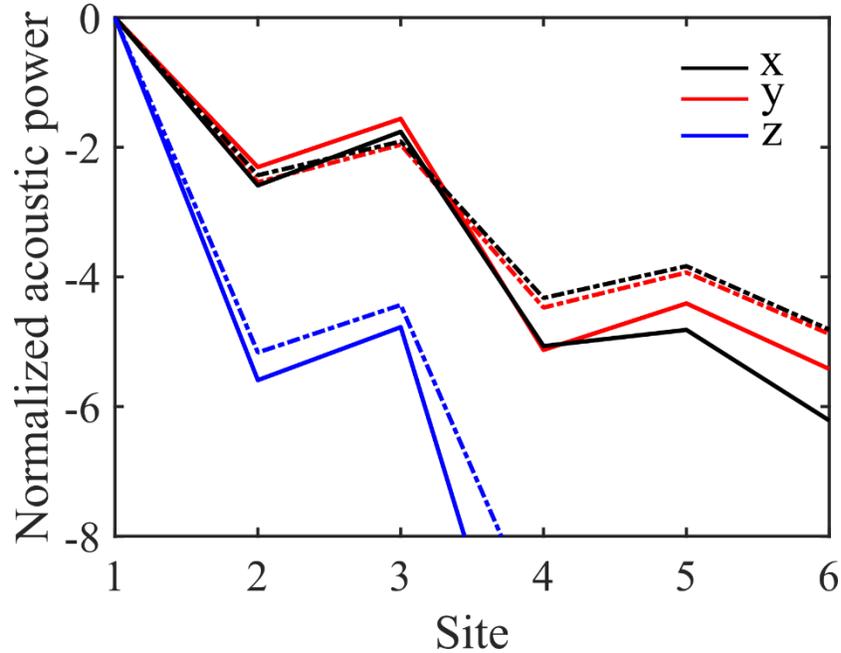

**Extended Data Figure 3 | Normalized acoustic field (in logarithmic scale) of corner mode excited at mid-gap frequency** distributed in the x hinge (black colored), y hinge (red colored) and z hinge (blue colored), respectively, the source is placed at one of corners (other corners show consistent behavior). Solid lines represent the results from first principle simulation with uniform loss introduced in the model, and dash-dotted lines represent the results from experiment. The average localization lengths agree well between theory and experiment and are found to be $0.45a_\perp$ and $0.52a_\perp$, respectively, along x/y directions, and $0.19a_z$ and $0.18a_z$, respectively, along z-direction. The localization length in z is shorter than that in x-y plane due to the intentionally chosen stronger dimerization of the inter-cell and intra-cell hopping in z direction than the ones in x-y plane. The experimental data shows larger decay of corner modes than what the theory predicts in x-y plane (clearly seen after the 3$^{rd}$ site), which is attributed to loss during the penetration along the hinge. In z-direction, the signal along the hinge after 3$^{rd}$ site is practically undetectable in the experiment due to stronger field localization. The strong localization of corner modes occurs due to strong dimerization (wide band gaps), which makes it possible to detect and characterize corner states even in a relatively small sample. It is also important to note that, instead of monotonically decreasing, both the experiment and simulations show oscillating behavior of the field distributions, which farther emphasizes the chiral symmetry of the system. The chiral symmetry enforces the sites that belong to the same sub-lattice as the corner site to carry larger weight of the wave function when compared to their neighbors from different sublattices.

**Data availability**

Data that are not already included in the paper and/or in the Supplementary Information are available on request from the authors.

**Acknowledgements**

The work was supported by the National Science Foundation with grants No. DMR-1809915, EFRI-1641069, and by the Defense Advanced Research Project Agency. X. N. thanks the support of Mina Rees Dissertation Fellowship at Graduate Center of CUNY.

**Author contributions**

All authors contributed extensively to the work presented in this paper.

**Author Information**

The authors declare no competing interests. Correspondence and requests for materials should be addressed to Andrea Alù or Alexander B. Khanikaev.